\documentclass[conference,compsoc]{IEEEtran}
\ifCLASSOPTIONcompsoc
  \usepackage[nocompress]{cite}
\else
  \usepackage{cite}
\fi
\ifCLASSINFOpdf
\else
\fi
\usepackage{amsmath,amssymb}
\usepackage{mathtools}
\usepackage{tikz}
\usetikzlibrary{arrows,fit,positioning}
\usepackage{pgfplots}
\pgfplotsset{width=10cm,compat=1.9}
\usepackage[caption=false,font=footnotesize]{subfig}
\usepackage[export]{adjustbox}
\usepackage{tabularx}
\usepackage{colortbl}
\usepackage{multirow}
\usepackage{stmaryrd}
\usepackage{arydshln}
\usepackage{amsmath,amssymb}
\usepackage{mathtools}
\usepackage{formal-grammar}
\usepackage{tikz}
\usetikzlibrary{arrows,fit,positioning,shapes,patterns}
\usepackage{pgfplots}
\usepackage{forest}
\usepackage{soul}
\usepackage{listings}
\pgfplotsset{width=8cm,compat=1.9}
\definecolor{scgreen}{HTML}{7fc97f}
\definecolor{scgreen2}{HTML}{b2df8a}
\definecolor{scorange}{HTML}{fdc086}
\definecolor{scpurple}{HTML}{beaed4}
\definecolor{scred}{HTML}{e41a1c}
\definecolor{scdgreen}{HTML}{4daf4a}
\definecolor{scblue}{HTML}{a6cee3}
\usepackage{algorithm}
\usepackage{algpseudocode}
\newcommand{\egg}{\texttt{egg}}
\newcommand{\assume}{\texttt{ASSUME}}
\newcommand{\reg}{\texttt{REG}}
\newcommand{\gate}{\texttt{TREG}}
\newcommand*\change[1]{#1}

\usepackage{booktabs}

\usepackage{soul}

\usepackage{tabularx}
\usepackage{colortbl}
\usepackage{multirow}
\usepackage{arydshln}

\usepackage[caption=false,font=footnotesize]{subfig}
\usepackage[export]{adjustbox}

\usepackage{soul}

\newcommand{\mycc}{\cellcolor{lightgray!50}}
\newcommand{\ccp}{\cellcolor{scgreen!50}}
\newcommand{\ccg}{\cellcolor{scpurple!50}}

\newcommand{\maxpowred}{33.9\%}
\newcommand{\avgpowred}{23.6\%}
\newcommand{\avgareachg}{5.0\%}

\newcommand{\rov}{ROVER}
\newcommand*\best[1]{{\bf #1}}

\begin{document}

\title{Combining Power and Arithmetic Optimization via Datapath Rewriting}

\author{\IEEEauthorblockN{Samuel Coward$^{1,2}$, Theo Drane$^1$, Emiliano Morini$^1$ and George A.~Constantinides$^2$}
\IEEEauthorblockA{$^1$ Intel Corporation, $^2$ Imperial College London,\\
Email: \{samuel.coward, theo.drane, emiliano.morini\}@intel.com, g.constantinides@imperial.ac.uk}
}
\maketitle

% \gc{Think about whether we can have a snappier title; `to reduce' sounds a little weak. Maybe something like `Datapath Power Optimization: A Rewriting Approach', or similar?}
\begin{abstract}
Industrial datapath designers consider dynamic power consumption to be a key metric. Arithmetic circuits contribute a major component of total chip power consumption and are therefore a common target for power optimization. While arithmetic circuit area and dynamic power consumption are often correlated, there is also a tradeoff to consider, as additional gates can be added to explicitly reduce arithmetic circuit activity and hence reduce power consumption. In this work, we consider two forms of power optimization and their interaction: circuit area reduction via arithmetic optimization, and the elimination of redundant computations using both data and clock gating. By encoding both these classes of optimization as local rewrites of expressions, our tool flow can simultaneously explore them, uncovering new opportunities for power saving through arithmetic rewrites using the e-graph data structure. Since power consumption is highly dependent upon the workload performed by the circuit, our tool flow facilitates a data dependent design paradigm, where an implementation is automatically tailored to particular contexts of data activity. 
We develop an automated RTL to RTL optimization framework, \rov, that takes circuit input stimuli and generates power-efficient architectures. We evaluate the effectiveness on both open-source arithmetic benchmarks and benchmarks derived from Intel production examples. The tool is able to reduce the total power consumption by up to \maxpowred. 
\end{abstract}


% \fi
\IEEEpeerreviewmaketitle

%%%%%%%%%%%%%%%%%%%%%%%%%%%%%%%%%%%%%%%%%%%%%%%%%%%%%%%%%%%%
% INTRODUCTION
%%%%%%%%%%%%%%%%%%%%%%%%%%%%%%%%%%%%%%%%%%%%%%%%%%%%%%%%%%%%
\section{Introduction}\label{sec:intro}
The three mostly common circuit quality metrics used in digital hardware design are power, performance and area, abbreviated to PPA. The performance of a circuit refers to how fast the circuit can execute the specified computation, the area is a measurement of how much silicon the circuit occupies on a die, which is highly correlated with manufacturing cost. Lastly, power is a measurement of the energy per unit time used to perform a given computation. The rise of custom accelerators presents the opportunity to optimize arithmetic hardware designs for particular computations, allowing us to perform those computations using less energy. 
% \gc{Two issues with original power phrasing: firstly it's at odds with the others, in that you focus on a unit (equivalent statement for area would be `it's how many square microns...'), secondly we can't talk of the `power used to perform a given computation', only of the `energy used to perform a given computation'... the physics doesn't make sense otherwise. Have suggested an alternative.} 
% Performance per watt is widely used to compare different computer chips. Such a figure can be improved by increasing performance within the same power budget or by reducing the power consumed to do the same computation \gcc{}{within the same time}. \gc{(Similar issue)} \gc{Here there's a focus on peformance in your opening discussion, yet performance doesn't feature heavily in the discussions we've had about this work to date... consider whether this framing is consistent with the results.}
\begin{figure}
    \centering
    \begin{tikzpicture}

	% Place nodes using a matrix
    \node at (0,0.7) (a)  {$A$};
    \node at (0,1.15) (b) {$B$};
    \node at (0,1.65) (c) {$C$};
    \node at (4,2.3) (s) {$S$};
    \node at (4,0.7) (one) {1};
    \node at (4,1.4) (zero) {0};
    \node at (5,1) (out) {out};

    \node[draw,shape=rectangle, minimum height=0.75cm] at (2.5,1.4) (mul) {$*$};
    % \node[draw,shape=rectangle, minimum height=0.75cm] at (1.5,1.5) (mul) {$\&$};
    \draw[red] (1,1.95) arc(90:-90:0.6cm and 0.3cm);
    \draw[red] (1,1.95) -> (1,1.35);
    \node[draw,shape=trapezium,inner xsep=16pt, inner ysep=6pt,rotate=270] at (4,1) (sel) {};
    \node[draw, red, circle, inner sep=1.5pt] at (0.9,1.8) (not) {};

    \draw[->] (a) -> (one);
    \draw[->] (b) -> (2.25,1.15);
    \draw[->] (c) -> (2.25,1.65);
    \draw[->] (mul) -> (zero);
    \draw[->] (s) -> (sel);
    \draw[->] (sel) -> (out);

    \draw[->,red] (s) -- (0.5,2.3) -- (0.5,1.8) -- (not);

 \end{tikzpicture}
    \caption{An operand isolation opportunity. In the original circuit (black), the input to the multiplier can be data gated when the select signal is one, as shown by the red gate. The negated select signal, $\overline{S}$ is a common input to an array of AND gates equal to the bitwidth of $C$.}
    \label{fig:op_isolate}
\end{figure}
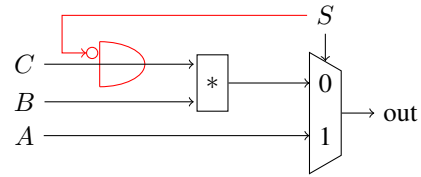 

Whilst performance and area are relatively simple to estimate, power can only be estimated accurately knowing the data values being operated upon, e.g. via representative workloads. The majority of power estimation tools in the electronic design automation (EDA) industry are based on randomly generated or real-world simulation stimuli~\cite{Synopsys2023PowerCompiler,Cadence2023JoulesSolution}. In this work, we focus on dynamic power consumption that is influenced heavily by register transfer level (RTL) design. From a simulation of a circuit design it is possible to infer bit-level switching activity, that is the frequency of transitions of a bit from zero to one or one to zero. 
These switching activities can be translated into power consumption estimates via a power model. 

Recently, Coward \textit{et al.} developed an RTL rewriting framework to address both area~\cite{Coward2022AutomaticE-Graphs} and performance~\cite{Coward2023AutomatingE-Graphs}. The core of their approach is an e(quivalence)-graph representation of RTL that facilitates exploration of these two metrics. In this work, we leverage this RTL rewriting framework and complete the PPA axes. We encode power optimizations such as clock gating and operand isolation via a set of local rewrites that, when combined with arithmetic rewrites, facilitate the exploration of new design spaces and the discovery of novel power efficient designs. 
We also exploit the compact e-graph representation to develop a computationally efficient power model that enables data-dependent circuit design. We develop a tool, \rov, which can customize an implementation based on representative workloads. 

In Section~\ref{sec:background} we provide background on RTL power analysis and optimization along with an introduction to e-graphs. In Section~\ref{sec:methodology} we describe how \rov~encodes power optimizations in terms of RTL rewrites. We also describe \rov's power model and how the compact representation provided by the e-graph yields efficiency gains for RTL simulation. Lastly, in Section~\ref{sec:results} we demonstrate \rov's impact on power consumption on a set of benchmarks.

The paper contains the following novel contributions:
\begin{itemize}
    \item a set of local equivalence preserving RTL rewrites that capture power-specific optimizations, 
    \item an encoding of clock gating and operand isolation that goes beyond current mux tree analysis,
    \item a computationally-efficient methodology to simulate a large set of design choices, leveraging the compact e-graph representation, 
    \item an automated method for data-driven design.
\end{itemize}

%%%%%%%%%%%%%%%%%%%%%%%%%%%%%%%%%%%%%%%%%%%%%%%%%%%%%%%%%%%%
% BACKGROUND
%%%%%%%%%%%%%%%%%%%%%%%%%%%%%%%%%%%%%%%%%%%%%%%%%%%%%%%%%%%%
\section{Background} \label{sec:background}
\subsection{RTL Power Optimization and Analysis}\label{subsec:opt_background}
Power optimizations can be broadly separated into two groups. First, a set of optimizations that primarily target circuit area reduction, since there is a correlation between circuit area and dynamic power consumption. This is intuitive because a smaller circuit area corresponds to fewer gates and thus fewer gates to toggle. Several previous works from both academia~\cite{dataflow2008verma,DeDinechin2021TowardsDesign,Ustun2022IMpress:HLS} and industry~\cite{Zimmermann2009DatapathDesign} have explored datapath RTL area reduction, including one work that used e-graph rewriting~\cite{Coward2022AutomaticE-Graphs}.

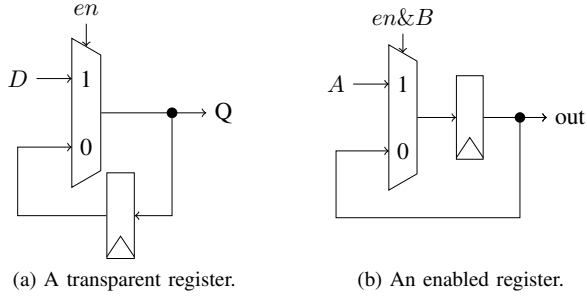
\begin{figure}
    \centering
    \subfloat[A transparent register.] {\resizebox{0.4\columnwidth}{!}{
\begin{tikzpicture}

	% Place nodes using a matrix
    \node at (3,1.5) (d)  {$D$};
    % \node at (0,1.25) (b) {$B$};
    % \node at (0,1.75) (c) {$C$};
    \node at (4,2.5) (en) {$en$};
    \node at (4,0.5) (zero) {0};
    \node at (4,1.5) (one) {1};
    \node [draw,circle, fill,inner sep=0.05cm] at (5.25,1) (fork) {};
    \node at (6,1) (out) {Q};

    % \node [draw,triangle, fill] at (5.75,1) (fork) {};

    \node[draw,shape=rectangle, minimum height=1.25cm, minimum width=0.4cm] at (4.5,-0.5) (reg) {};
    \node[draw,shape=trapezium,inner xsep=24pt, inner ysep=6pt,rotate=270] at (4,1) (sel) {};

    \draw[->] (d) -> (one);
    \draw[->] (en) -> (sel);
    % \draw[->] (reg) -> (out);
    % \draw[->] (sel) -> (reg);
    \draw[->] (sel) -> (out);

    \draw[->] (fork) |- (reg) ;
    \draw[->] (reg) -- (3,-0.5) -- (3,0.5) -- (zero);
    \draw[-] (4.3,-1.1) -- (4.5,-0.8) -- (4.7,-1.1);

 \end{tikzpicture}
 }\label{fig:transparent}}
    \qquad
    \subfloat[An enabled register.] {\resizebox{0.45\columnwidth}{!}{
\begin{tikzpicture}

	% Place nodes using a matrix
    \node at (3,1.5) (d)  {$A$};
    % \node at (0,1.25) (b) {$B$};
    % \node at (0,1.75) (c) {$C$};
    \node at (4,2.5) (en) {$en\& B$};
    \node at (4,0.5) (zero) {0};
    \node at (4,1.5) (one) {1};
    \node [draw,circle, fill,inner sep=0.05cm] at (5.75,1) (fork) {};
    \node at (6.5,1) (out) {out};
    \node at (4,-1) {};

    % \node [draw,triangle, fill] at (5.75,1) (fork) {};

    \node[draw,shape=rectangle, minimum height=1.25cm, minimum width=0.4cm] at (5,1) (reg) {};
    \node[draw,shape=trapezium,inner xsep=24pt, inner ysep=6pt,rotate=270] at (4,1) (sel) {};

    \draw[->] (d) -> (one);
    % \draw[->] (b) -> (2.25,1.25);
    % \draw[->] (c) -> (2.25,1.75);
    % \draw[->] (mul) -> (zero);
    \draw[->] (en) -> (sel);
    \draw[->] (reg) -> (out);
    \draw[->] (sel) -> (reg);
    \draw[->] (reg) -> (out);

    \draw[->] (fork) -- (5.75,-0.5) -- (3,-0.5) -- (3,0.5) -- (zero);
    \draw[-] (4.8,0.4) -- (5,0.7) -- (5.2,0.4);

 \end{tikzpicture}
 }\label{fig:register}}
    \caption{Circuit diagrams of the \gate~and \reg~operators.}
    \label{fig:circuits}
\end{figure}

The second set of optimizations, which are the primary focus of this work, detects opportunities to switch off, or gate, sub-circuits in the design. Clock gating and operand isolation are two such optimizations. For a clock gating example, consider a pipelined floating-point adder in which exception cases, {\em e.g.}~NaNs, are handled on a separate exception path. If we detect an exceptional input in the first stage, we can gate all registers on the standard input path for subsequent stages, since the result is redundant. Gating the registers stops the register outputs changing and hence prevents any toggling of the downstream combinational logic. The additional gating logic adds an area (and possible delay) overhead which must be evaluated alongside the data-dependent power saving. For an operand isolation example, consider Figure~\ref{fig:op_isolate}. In this circuit we can identify a redundant operation and construct an activation signal that we use to zero one of the multiplier inputs, limiting operator power consumption. We refer to this technique as data gating. Alternatively, both multiplier inputs could be ``frozen'' using transparent registers to eliminate redundant toggling~\cite{Munch2000AutomatingDatapaths,Tiwari1998GuardedSynthesis/design}. The transparent register shown in Figure~\ref{fig:transparent} has an enable signal, which, when high, allows the input to transparently flow through to the output and, when low, freezes the output. Prior work has called this circuit a transparent latch, but since we operate in a synchronous domain we shall instead call it a transparent register.

In academia, clock gating has been explored at a gate-level~\cite{Hurst2008AutomaticPerturbation} and from a clock tree synthesis perspective~\cite{Donno2003Clock-treeClock-gating}. A subset of industrial tools, such as Synopsys Power Compiler~\cite{Synopsys2023PowerCompiler} and Cadence Joules~\cite{Cadence2023JoulesSolution}, are incorporated into the logic synthesis engines and automatically perform clock-gating optimizations. Siemens PowerPro~\cite{PowerPro2021} is a standalone RTL to RTL tool that targets sequential clock gating. A limitation of these approaches is that they rely on analyzing the mux tree structure of the RTL design, but this may miss opportunities as we shall see in Section~\ref{subsec:gating}. The automation of operand isolation has been explored at both the word-level~\cite{Munch2000AutomatingDatapaths} and at gate-level~\cite{Tiwari1998GuardedSynthesis/design,Hoang2012Data-width-drivenCircuits}. An e-graph based rewriting approach allows us to compare and combine these different power optimization techniques.

RTL power analysis tools typically rely on simulation to estimate power consumption of a given design. Tool users can either provide simulation stimuli or set input switching activities and static probabilities~\cite{Synopsys2023PowerCompiler,Cadence2023JoulesSolution}. For a given simulation period, the switching activity describes how frequently each bit of the given signal transitions from zero to one or vice versa, and the static probability specifies what proportion of the time that bit is expected to be in the one state. Commercial logic synthesis tools~\cite{Synopsys2021DesignS-2021.06-SP2}, take user provided simulation configurations and perform power optimizations guided by the simulation. 

This paper describes how to encode the power optimizations discussed above as local RTL rewrites and a framework to explore the combination of power, arithmetic and area optimizations. Our tool encodes arithmetic optimizations performed by downstream synthesis tools in the optimization flow, a key factor not considered by prior work. For analysis the underlying data structure of our tool also enables efficient evaluation of the power consumption of equivalent design candidates.

%%%%%%%%%%%%%%%%%%%%%%%%%%%%%%%%%%%%%%%%%%%%%%%%%%%%%%%%%%%%
% EGRAPHS
%%%%%%%%%%%%%%%%%%%%%%%%%%%%%%%%%%%%%%%%%%%%%%%%%%%%%%%%%%%%
\subsection{E-Graphs}\label{subsec:egraphs_background}

\begin{figure}
    \centering
    \subfloat[Initial e-graph contains $A\, \& \,\{8\{M\}\}$] {\includegraphics[scale=0.3]{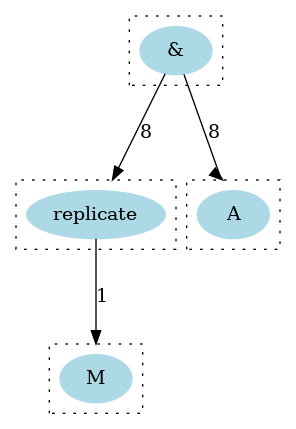}}
    \qquad
    \subfloat[Apply rewrite $A \& \{8\{M\}\} \rightarrow \gate(A,M) \& \{8\{M\}\}$] {\includegraphics[scale=0.3]{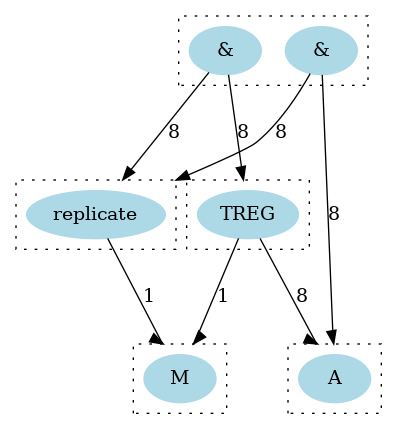}}
    \caption{E-graph rewriting of a masking operation. Dashed boxes represent e-classes of equivalent expressions. A new equivalent expression is added to the e-graph represented by the second $\&$ operator in the root e-class.}
    \label{fig: e-graph_example}
\end{figure}

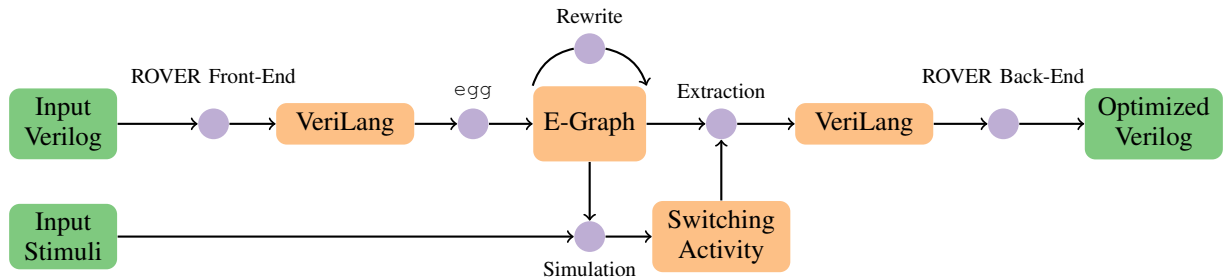
\begin{figure*}
    \centering
    \begin{tikzpicture}

% Boxes
\node [shape=rectangle, fill=scgreen, text width = 1.2cm, text centered,rounded corners] at (-5.5,0) (input) {Input Verilog};
% \node [shape=circle, fill=scpurple, text width = 0.1cm, text centered,label={\footnotesize{\texttt{Slang}}}] at (-6.5,0) (slang) {};
% \node [shape=rectangle, fill=scgreen, text width = 1.2cm, text centered,rounded corners] at (-5,0) (json) {JSON};
\node [shape=circle, fill=scpurple, text width = 0.1cm, text centered,label={[label distance=0.2cm]\footnotesize{\rov~Front-End}}] at (-3.5,0) (front_end) {};
\node [shape=rectangle, fill=scorange, text width = 1.6cm, text centered,rounded corners] at (-1.75,0) (verilang) {VeriLang};

\node [shape=circle, fill=scpurple, text width = 0.1cm, text centered,label={\footnotesize{\egg}}] at (-0.05,0) (egg) {};
\node [shape=rectangle, fill=scorange, minimum width=1.5cm, minimum height=1cm,rounded corners] at (1.5,0) (egraph) {E-Graph};
\node [shape=circle, fill=scpurple, text width = 0.1cm, text centered,label=270:{\footnotesize{Simulation}}] at (1.5,-1.5) (sim) {};
\node [shape=rectangle, fill=scorange, text width = 1.6cm, text centered, rounded corners] at (3.25,-1.5) (toggle) {Switching Activity};

\node [shape=rectangle, fill=scgreen, text width = 1.2cm, text centered,rounded corners] at (-5.5,-1.5) (data) {Input Stimuli};

\node [shape=circle, fill=scpurple, text width = 0.1cm, text centered,label={\footnotesize{Rewrite}}] at (1.5,1) (rewrites) {};
\node [shape=circle, fill=scpurple, text width = 0.1cm, text centered,label={\footnotesize{Extraction}}] at (3.25,0) (extract) {};
\node [shape=rectangle, fill=scorange, text width = 1.6cm, text centered, rounded corners] at (5.15,0) (opt_verilang) {VeriLang};
\node [shape=circle, fill=scpurple, text width = 0.1cm, text centered,label={[label distance=0.2cm]\footnotesize{\rov~Back-End}}] at (7,0) (back_end) {};
\node [shape=rectangle, fill=scgreen, text width = 1.6cm, text centered, rounded corners] at (9,0) (output) {Optimized Verilog};

% \node [shape=circle, fill=scpurple, text width = 0.1cm, text centered,label=0:{\footnotesize{Proof Production}}] at (5.15,-1) (proof_prod) {};

% \node [shape=rectangle, fill=scorange, text width = 1.6cm, text centered, rounded corners] at (5.15,-2) (proof) {VeriLang};

% \node [shape=circle, fill=scpurple, text width = 0.1cm, text centered,label={[label distance=0.05cm]30:\footnotesize{\rov~Back End}}] at (5.15,-3) (back_end_proof) {};
% \node [shape=rectangle, fill=scgreen, text width = 2cm, text centered,rounded corners] at (-5,-4) (int1) {Intermediate Verilog 1};
% \node [shape=rectangle, fill=scgreen, text width = 2cm, text centered,rounded corners] at (-1,-4) (int2) {Intermediate Verilog 2};
% \node [shape=ellipse] at (2.5,-4) (dots) {$\cdot \cdot \cdot$};
% \node [shape=rectangle, fill=scgreen, text width = 2cm, text centered,rounded corners] at (7,-4) (intn) {Intermediate Verilog $n$};

% \node [shape=circle, fill=scpurple, text width = 0.1cm, text centered,label=45:{\footnotesize{EC}}] at (-8,-4) (fv_0) {};

% \node [shape=circle, fill=scpurple, text width = 0.1cm, text centered,label={\footnotesize{EC}}] at (-3,-4) (fv_1) {};

% \node [shape=circle, fill=scpurple, text width = 0.1cm, text centered,label=135:{\footnotesize{EC}}] at (9,-4) (fv_2) {};

% \draw [->,thick] (input) edge (slang);
% \draw [->, thick] (slang) edge (json);
\draw [->, thick] (input) edge (front_end);
\draw [->, thick] (front_end) edge (verilang);
\draw [->, thick] (verilang) edge (egg);
\draw [->, thick] (egg) edge (egraph);
\path [->, thick] (rewrites.east) edge[bend left] (egraph.north east);
\draw [-, thick] (egraph.north west) edge[bend left] (rewrites.west);

\draw [->, thick] (egraph) edge (extract);
\draw [->, thick] (egraph) edge (sim);
\draw [->, thick] (data) edge (sim);
\draw [->, thick] (sim) edge (toggle);
\draw [->, thick] (toggle) edge (extract);
\draw [->, thick] (extract) edge (opt_verilang);
\draw [->, thick] (opt_verilang) edge (back_end);
\draw [->, thick] (back_end) edge (output);

% \draw [->,  thick] (opt_verilang) edge (proof_prod);
% \draw [->,  thick] (proof_prod) edge (proof);
% \draw [->,  thick] (proof) edge (back_end_proof);
% \draw [->,  thick] (back_end_proof) -| (int1);
% \draw [->,  thick] (back_end_proof) -| (int2);
% \draw [->,  thick] (back_end_proof) -| (intn);
% \draw[->,  thick] (input) edge (fv_0);
% \draw[->,  thick] (int1) edge (fv_0);
% \draw[->,  thick] (int1) edge (fv_1);
% \draw[->,  thick] (int2) edge (fv_1);

% \draw[->,  thick] (intn) edge (fv_2);
% \draw[->,  thick] (output) edge (fv_2);

\end{tikzpicture}
    \caption{\rov's power optimization tool flow. Users provide an input Verilog design and input stimuli via simulation data or switching activity statistics.}
    \label{fig:flow-diagram}
\end{figure*}

An e-graph is typically initialized with a single expression, such that every e-class contains a single node. The e-graph is grown via constructive rewrite application, where a pattern, $l$, is matched in the e-graph and gets rewritten to a different expression, $r$. The e-graph retains the original expression $l$ in the data structure, $r$ is simply added to the matched e-class. Figure~\ref{fig: e-graph_example} shows an example e-graph before and after applying a rewrite. Given a set of rewrites, all rewrites are applied to the e-graph at each rewriting iteration. Determining an optimal sequence of rewrite applications is deferred to the extraction phase. We grow an e-graph until we reach a computational limit or we reach a state called saturation, where further rewriting adds no new nodes to the data structure. The final e-graph represents a set of equivalent implementations. The process to select the `best' implementation is known as extraction and is typically based on a custom cost model for each application. 

Coward, {\em et al.}~introduced an intermediate language, VeriLang, that can represent combinational RTL in an e-graph~\cite{Coward2022AutomaticE-Graphs,Coward2023AutomatingE-Graphs}. We adopt the same representation here, but extend it to incorporate registers, allowing us to represent pipelined designs in the e-graph. 
Earlier work by the same authors
%of~\cite{Coward2022AutomaticE-Graphs} and~\cite{Coward2023AutomatingE-Graphs} 
developed an RTL optimization tool using mixed precision RTL rewriting to reduce circuit area and delay. We build on their framework as a foundation, extending it to target power optimization. Our key insight is that their delay optimization work~\cite{Coward2023AutomatingE-Graphs}, introduced \assume~nodes to exploit the mux tree structure of a design in an e-graph rewriting framework. The \assume~node associates code branches with \textit{observability don't care} conditions and allow the e-graph to capture equivalence under some constraint, which they show to be useful in identifying logic to accelerate special cases, e.g. near/far path floating point addition. As noted above, \textit{observability don't cares} are also -- for other reasons -- of central relevance to power-saving optimizations in RTL datapath design, and we describe in this paper how to exploit that commonality.

Recently, a general purpose and extensible e-graph library, \egg, was released~\cite{Willsey2021Egg:Saturation}. \egg~provides a number of innovations in performance and e-graph features that has led to applications in numerical stability improvement~\cite{Panchekha2015AutomaticallyExpressions} and FPGA multiplier design~\cite{Ustun2022IMpress:HLS}.

%%%%%%%%%%%%%%%%%%%%%%%%%%%%%%%%%%%%%%%%%%%%%%%%%%%%%%%%%%%%
% METHODOLOGY
%%%%%%%%%%%%%%%%%%%%%%%%%%%%%%%%%%%%%%%%%%%%%%%%%%%%%%%%%%%%
\section{Methodology}\label{sec:methodology}
In Figure~\ref{fig:flow-diagram} we provide an overview of how \rov~optimizes a design to reduce power consumption. Using the \rov~front-end, an input Verilog design is converted to VeriLang, which \egg~uses to initialize an e-graph~\cite{Coward2022AutomaticE-Graphs}. \rov~then applies a set of power optimization rewrites, described in Sections~\ref{subsec:rewriting} and \ref{subsec:gating}, to the e-graph, constructing a set of implementation candidates. To accurately model per implementation power consumption, \rov~simulates the entire e-graph based on user configured input stimuli as described in Section~\ref{subsec:simulation}. The user configured input stimuli provide, for every module input, a sequence of bitvectors that are fed one per clock cycle. Throughout this work, we assume a single clock domain, for simplicity. E-graph simulation provides switching activities for all the internal signals of all the candidates, which are fed into the power model, described in Section~\ref{subsec:op_model}. The power model is used by \rov~to determine the optimal implementation, producing a VeriLang expression. The \rov~back-end converts the extracted VeriLang expression into Verilog. 

In the original paper where VeriLang was introduced~\cite{Coward2022AutomaticE-Graphs}, the authors envisaged its semantics as operating over Boolean values. In this work, we modify the semantics of VeriLang to consider input variables as {\em streams} of Boolean data, such that a new data point enters the module every clock cycle. Since we consider streams of data, every intermediate signal created in the e-graph has an associated stream. The semantics of combinational operators are such that each clock cycle the new data points are used to generate a new output within the same cycle. 

We incorporate a new VeriLang operator, \reg, which describes a register with an enable signal. The corresponding  circuit is shown in Figure~\ref{fig:register}. Given input $a$ and enable signal $en$ with associated data streams $a_i$ and $en_i$, $\reg(a,en)$ has the following semantics:
\begin{equation}\label{eqn:reg_semantics}
    \reg(a_i,en_i) =
    \begin{cases}
        0 &\text{, if }i==0\\
        a_{i-1} &\text{, if }en_{i-1}\\
        \reg(a_{i-1},en_{i-1}) &\text{, else.} \\
        
    \end{cases}
\end{equation}
These semantics assume that a register is initialized to zero.

%%%%%%%%%%%%%%%%%%%%%%%%%%%%%%%%%%%%%%%%%%%%%%%%%%%%%%%%%%%%
% DATA GATING
%%%%%%%%%%%%%%%%%%%%%%%%%%%%%%%%%%%%%%%%%%%%%%%%%%%%%%%%%%%%
\subsection{Data Gating} \label{subsec:rewriting}
\begin{table*}
    \centering
    \setlength\extrarowheight{2.5pt}
    \caption{
    A set of RTL rewrites encoding operand isolation and clock gating optimizations. 
    We define four sets of operators such that $\texttt{op}$ is any arithmetic or logical VeriLang operator, $\texttt{op1} \in \{*, \ll, \gg, +,-\}$, $\texttt{op2} \in \texttt{op1}\setminus\{+,-\}$ and $\texttt{op3}$ is any Boolean operator. We use $w_a$ to denote the bitwidth of a bitvector $a$, $w_o$ to denote the output bitwidth of an operation.
    }
    \begin{tabular}{llll}
    \toprule
    Group & Name      & Left-Hand Side                        & Right-Hand Side \\
    \midrule
    \multirow{8}{5em}{Data Gate} 
    & \mycc Gate Left           & \mycc $s\,?\,b\,:\,c$                       
                                & \mycc $s\,?\,\left(b\,\&\,\{w_b\{ s\}\}\right)\,:\,c$\\
    & Gate Right                & $s\,?\,b\,:\,c$                       
                                & $s\,?\,b\,:\,\left(c\,\&\,\{w_c\{ \overline{s}\}\}\right)$\\
    & \mycc Propagate Mask      & \mycc $(a\; \texttt{op1}\; b)\,\& \, \{w_o\{s\}\}$ 
                                & \mycc$(a\,\& \, \{w_a\{s\}\})\; \texttt{op1}\; (b\,\& \, \{w_b\{s\}\})$ \\
    & Propagate Mask Left       & $(a\; \texttt{op2}\; b)\,\& \, \{w_o\{s\}\}$ 
                                & $(a\,\& \, \{w_a\{s\}\})\; \texttt{op2}\; b$ \\
    & \mycc Propagate Mux Mask  & \mycc $(s_1\, ?\, a\,:\, b)\,\& \, \{w_o\{s_2\}\}$ 
                                & \mycc $s_1\, ?\, (a\,\& \, \{w_a\{s_2\}\})\,:\, (b\,\& \, \{w_b\{s_2\}\})$ \\
    & Propagate Mux Mask Right  & $(s_1\, ?\, a\,:\, b)\,\& \, \{w_o\{s_2\}\}$ 
                                & $s_1 \& s_2 \, ?\, a\,:\, (b\,\& \, \{w_b\{s_2\}\})$ \\
    & \mycc Propagate Mux Mask Left   & \mycc $(s_1\, ?\, a\,:\, b)\,\& \, \{w_o\{s_2\}\}$ 
                                & \mycc $s_1 \,\|\, \overline{s_2} \, ?\, (a\,\& \, \{w_a\{s_2\}\})\,:\, b$ \\
    & Combine Masks       & $(a\,\& \, \{w_a\{s_1\}\}) \,\& \, \{w_a\{s_2\}\}$          
                                & $a\,\& \, \{w_a\{s_1\& s_2\}\}$ \\
    \midrule
    \multirow{8}{5em}{Transparent Registers} 
    % & Transp Reg Branches     & $s\,?\,b\,:\,c$                 
                                % & $s\,?\,\gate{(b,s)}\,:\,\gate(c,\overline{s})$\\
    & \mycc Transp Reg Left   & \mycc $s\,?\,b\,:\,c$                 
                                & \mycc $s\,?\,\gate{(b,s)}\,:\,c$                     \\
    & Transp Reg Right        & $s\,?\,b\,:\,c$                 
                                & $s\,?\,b\,:\,\gate(c,\overline{s})$            \\
    & \mycc Transp Reg Mask   & \mycc $a\,\&\,\{w_a\{ s\}\}$                 
                                & \mycc $\gate(a,s)\,\&\,\{w_a\{ s\}\}$            \\
    & Transp Reg Saturate     & $a\,\|\,\{w_a\{ s\}\}$                 
                                & $\gate(a,\overline{s})\,\|\,\{w_a\{ s\}\}$            \\
    & \mycc Transp Reg Reg    & \mycc $\reg(a,en)$                 
                                & \mycc $\reg\left(\gate(a,en),en\right)$            \\
    & Propagate                 & $\gate(a\; \texttt{op}\; b,s)$  
                                & $\gate(a,s)\; \texttt{op}\; \gate(b,s)$ \\
    & \mycc Propagate Mux       & \mycc $\gate(s_1\, ?\, a\,:\, b,s_2)$ 
                                & \mycc $\gate(s_1,s_2)\, ?\, \gate(a,s_2)\,:\, \gate(b,s_2)$ \\
    % & Propagate Mux Right       & $\gate(s_1\, ?\, a\,:\, b,s_2)$ 
                                % & $s_1\& s_2\, ?\, a\,:\, \gate(b,s_2)$ \\
    % & \mycc Propagate Mux Left  & \mycc $\gate(s_1\, ?\, a\,:\, b,s_2)$ 
                                % & \mycc $s_1\,\|\, \overline{s_2}\, ?\, \gate(a,s_2)\,:\, b$ \\
    & Combine Transp Reg      & $\gate(\gate(a,s_1),s_2)$       
                                & $\gate(a,s_1\,\&\,s_2)$ \\
    \midrule
    
    \multirow{2}{5em}{Clock Gate \& Retime} 
        &\mycc Retime Boolean & \mycc $\reg(a,en) \texttt{ op3 } \reg(b,en)$ 
                          & \mycc $\reg(a \texttt{ op3 } b, en)$\\
    & Clock Gate Reg            &$\gate(\reg(a, en), \reg(b,en))$ 
                                & $\reg(a,en \, \&\, b)$\\
    \bottomrule
    \end{tabular}
    \label{tab:rewrites}
\end{table*}
In this section we describe a set of rewrites that encodes the operand isolation optimizations described in Section~\ref{subsec:opt_background}. A key challenge in expressing operand isolation via local rewrites, is that having identified a redundant computation from a functional perspective we do not care what value is produced under certain conditions. Therefore, from a functional perspective, when we do not care, we can generate any value we choose. However, the values that we select have a significant impact upon power consumption. 

Before progressing, we define notation used throughout. We let $w_x$ denote the bitwidth of a bitvector variable $x$ and let $w_o$ denote the output bitwidth of an operation. We use $\{w \{ S \} \}$ to denote $w$-fold replication of a bitvector $S$, which will usually be a single bit. Lastly, we use $\overline{S}$ to denote the bitwise logical complement of a bitvector $S$.

In one approach to perform operand isolation we can apply data gating to each branch of a mux operator. Data gating creates a mask by duplicating a select signal and applies a bitwise AND operation. This rewrite explicitly zeroes redundant outputs. The first group in Table~\ref{tab:rewrites} contains rewrites to create initial data gating operations. We include two ``Gate'' rewrites as we may wish to data gate only the true branch, only the false branch or both, by applying ``Gate Left'' and ``Gate Right'' in sequence. The rewrite from \eqref{eqn:ex_initial} to \eqref{eqn:ex_gate} illustrates the creation of a mask and data gating of a mux branch, in order to avoid dynamic power in the multiplier. Table~\ref{tab:rewrites} next describes how the data gating operations are propagated over arithmetic operations, since these operators typically account for the largest power consumption in datapath circuits. The ``Propagate'' rewrites incrementally gate larger sub-circuits. For a subset of operators, {\em e.g.}~multiplication, it is equivalent to data gate a single operand, as 
illustrated in \eqref{eqn:ex_gate} and \eqref{eqn:ex_final}. The rewrites in Table~\ref{tab:rewrites} encode the optimization shown in Figure~\ref{fig:op_isolate}.
\begin{align}
    &S \; ? \; A \,: \, (C*B)\hspace{1.75cm} \rightarrow \textit{
    (Gate Right)} \label{eqn:ex_initial}\\
    &S \; ? \; A \,: \,(C*B)\,\&\,\{w_o\{\overline{S}\}\}\rightarrow \textit{(Propagate Left)} \label{eqn:ex_gate}\\
    &S \; ? \; A \,: \, (C\,\&\,\{w_c\{\overline{S}\}\})*B \label{eqn:ex_final}
\end{align}

The impact of data gating redundant operations depends on the wider module context. Exploring data gating via e-graph rewriting allows \rov~to retain a set of gated and ungated designs, deferring architecture selection and evaluation to the extraction phase. For example, in
\begin{equation}
    (s\, ?\, f(a)\, :\, b) + g(f(a)),
\end{equation}
the computation of $f(a)$ appears redundant when $s$ is zero, however we always use $f(a)$ in the computation of $g(f(a))$, thus there is no value in a gated version. 

Applying these rewrites to a nested mux structure, \rov~naturally generates nested gating operations which are combined via classical Boolean rewriting. Such an approach constructs \textit{observability don't care} conditions that are not present in the original design. These newly created conditions can be simplified using Boolean rewriting. 

In addition to the rewrites described in Table~\ref{tab:rewrites}, \rov~deploys the arithmetic and area optimization rewrites described in~\cite{Coward2022AutomaticE-Graphs}, that crucially encode downstream logic synthesis optimizations. \rov~also includes standard Boolean rewrites for optimizing logical expressions. Exploring these transformations in parallel, \rov~discovers architectures providing an efficient area power tradeoff.

%%%%%%%%%%%%%%%%%%%%%%%%%%%%%%%%%%%%%%%%%%%%%%%%%%%%%%%%%%%%
% CLOCK GATING
%%%%%%%%%%%%%%%%%%%%%%%%%%%%%%%%%%%%%%%%%%%%%%%%%%%%%%%%%%%%
\subsection{Clock Gating} \label{subsec:gating}
In the previous section, we described how data gating rewrites can encode operand isolation. In this section, we shall describe a set of local equivalence preserving rewrites that create transparent registers providing an alternative way to achieve operand isolation. In \eqref{eqn:reg_semantics} we defined the semantics of \reg. We define an additional VeriLang operator, \gate, representing a transparent register, shown in Figure~\ref{fig:transparent}, with semantics that are similar to those of \reg. 
\begin{equation*}\label{eqn:gate_semantics}
    \gate(a_i,b_i) =
    \begin{cases}
        a_i &\text{, if }b_i\\
        \gate(a_{i-1},b_{i-1}) &\text{, elif } i>0\\
        0 &\text{, else}
    \end{cases}
\end{equation*}

The second group in Table~\ref{tab:rewrites} contains a set of rewrites, similar to the first group, that encode the creation and propagation of \gate~operators. We improve upon approaches based on mux tree analysis by including the ``Transp Reg Mask/Saturate'' rewrites that detect redundant computation, as used in Figure~\ref{fig: e-graph_example}. We also create transparent registers from register enable signals, since disabled registers correspond to redundant computation. The ``Combine Transp Reg'' rewrite allows \rov~to construct complex \textit{observability don't care} signals that may not be present in the initial design. These signals may be simplified via Boolean rewriting.

In the final group in Table~\ref{tab:rewrites}, we describe how \rov~encodes clock gating via local rewrites. When the \gate~operator meets the output of a register, it represents an opportunity to refine the enable condition of the register, eliminating the overhead of the transparent register. We can prove the equivalence of 
\begin{align*}
   L_i & =\gate(\reg(a_i,en_i), \reg(b_i,en_i)) \text{ and }\\
   R_i & = \reg(a_i, en_i \& b_i)
\end{align*}
for all clock cycles $i$ via induction. First, let 
\[p_i=\reg(a_i,en_i) \text{ and }q_i=\reg(b_i,en_i).\]
Suppose $\forall i\leq k\;L_i = R_i$, then if $en_k=1$:
\begin{align*}
    q_{k+1} & = b_k \hspace{2em} p_{k+1} = a_k  \\
    L_{k+1} & = q_{k+1}\, ?\, p_{k+1}\, :\, L_k \\
            & = b_k\, \hspace{1.1em}?\, a_k\, \hspace{1em}:\, L_k\\
\end{align*}
Then, since $en_k=1$ and $R_k=L_k$, 
\begin{align*}
    R_{k+1} &= en_k \& b_k \, ? \, a_k \, : \, R_k \\ 
            &= \hspace{2.5em}b_k \, ? \, a_k \, : \, R_k = L_{k+1}
\end{align*}
Now if $en_k=0$, then $R_{k+1} = R_k=L_k$ and
\begin{align*}
    q_{k+1} & = q_k \hspace{2em} p_{k+1} = p_k  \\
    L_{k+1} & = q_k\, ?\, p_k\, :\, L_k \\
    q_k=1    \Rightarrow L_k&= q_k\,?\,p_k\,:\,L_{k-1}=p_k
\end{align*}
Therefore $L_{k+1}=L_k=R_{k+1}$ and hence $R_{k+1}=L_{k+1}$ for all values of $en_k$. Under the zero register initialization assumption it is trivial to prove $L_0=R_0$.

A key requirement of the ``Clock Gate Reg'' rewrite, is that the \textit{observability don't care} condition be available in the previous clock cycle. This constraint ensures that the register is disabled for the clock cycle corresponding to the redundant computation. In certain cases, it may be necessary to move operations into earlier clock cycles to ensure the gating signal is available in the correct cycle. To transfer operations between clock cycles we implement limited retiming of Boolean operators. 

As described in Figure~\ref{fig:flow-diagram}, \rov~applies all rewrites described to grow an e-graph of equivalent implementations until a user defined limit or saturation is reached. The final e-graph contains designs with different combinations of gating and arithmetic optimizations. Determining which combination of optimizations produce the most power efficient design is left to extraction, which we describe next.

%%%%%%%%%%%%%%%%%%%%%%%%%%%%%%%%%%%%%%%%%%%%%%%%%%%%%%%%%%%%
% SIMULATION
%%%%%%%%%%%%%%%%%%%%%%%%%%%%%%%%%%%%%%%%%%%%%%%%%%%%%%%%%%%%
\subsection{Simulation} \label{subsec:simulation}
In order to analyze power consumption, \rov~must first simulate all designs within the e-graph based on a set of stimuli. \rov~takes an additional input configuration file that provides simulation stimuli or sets the switching activity for each module input. If the user only defines a switching activity and simulation length, \rov~automatically generates simulation stimuli for all module inputs using \change{an algorithm that randomly toggles each bit in a bitvector according to the configured toggle rate.}

\begin{figure}
    \centering
    \resizebox{0.85\columnwidth}{!}{
\begin{tikzpicture}
\begin{axis}[
    xlabel={Number of E-Classes},
    ylabel={Number of Designs},
    width=\columnwidth,
    % xmin=0, xmax=1.05,
    % ymode=log,
    % ymin=0, ymax=120,
    % xtick={0,20,40,60,80,100},
    % ytick={0,20,40,60,80,100,120},
    % legend pos=north west,
    % ymajorgrids=true,
    % grid style=dashed,
]

% based on dg_weight rewriting
\addplot+[
    only marks,
    mark=square,
    color=red
    ]
    coordinates {
(11,1)
(20,15)
(34,137)
(42,546)
(38,942)
(38,1050)
};
\draw[->](axis cs:11.5,1.5)--(axis cs:19.5,14.5);
\draw[->](axis cs:20.5,15.5)--(axis cs:33.5,136.5);
\draw[->](axis cs:34.5,137)--(axis cs:42,526);
\draw[->](axis cs:42,566)--(axis cs:38,920);
\draw[->](axis cs:38,962)--(axis cs:38,1030);
\end{axis}

\end{tikzpicture}
}
    \caption{The number of designs vs. the number of e-classes after each iteration of rewriting the design in Figure~\ref{fig:op_isolate}. Simulation complexity scales with the number of e-classes but evaluates all designs in the e-graph.}
    \label{fig:expression_vs_classes}
\end{figure}
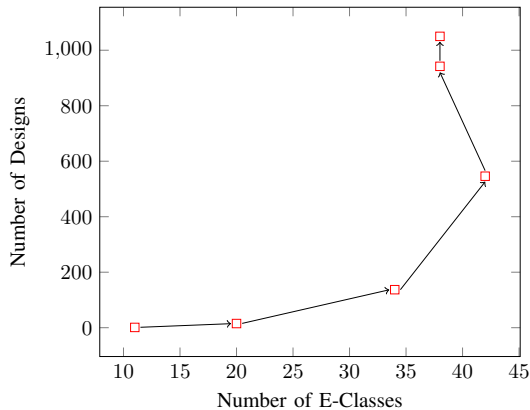

Since all nodes in a given e-class are functionally equivalent, \rov~simulates one node per e-class to obtain simulation data for the entire class. This observation provides a significant computational efficiency gain, as the complexity of simulating all designs in the e-graph scales with the number of e-classes. Meanwhile, the number of distinct designs contained in the e-graph can grow exponentially with the number of classes~\cite{Ustun2022IMpress:HLS}, as shown for the example of Figure~\ref{fig:op_isolate} in Figure~\ref{fig:expression_vs_classes}. In this example, the number of e-classes grows by a factor of four whilst the number of designs grows by a factor of 1000. The number of e-classes in Figure~\ref{fig:expression_vs_classes} does not grow monotonically, since in later rewriting iterations e-classes get merged due to proof of equivalence generated by \rov~reducing the number of classes. Note that, whilst a single node evaluation can be shared across the e-class, each node in the e-class may require more or less power to produce that same value. For example, $x+x$ and $x\ll 1$ are functionally equivalent but may consume significantly different power. It is this difference our extraction is designed to estimate and exploit.

From the e-class simulation data, \rov~calculates an average switching activity across the entire output word of that e-class. For example, for a 3-bit word with switching activities of 0.25 for bit 0, 0.5 for bit 1 and 0.75 for bit 2 would average to 0.5 across the entire word. 

%\begin{figure}
%    \centering
%    \input{gen_sim}
%    \caption{An algorithm to generate simulation stimuli of length $L$ for an individual bit at a given $\texttt{switching\_activity}\in[0,1]$.}\label{alg:gen_sim}
%\end{figure}

%%%%%%%%%%%%%%%%%%%%%%%%%%%%%%%%%%%%%%%%%%%%%%%%%%%%%%%%%%%%
% OPERATOR POWER MODEL
%%%%%%%%%%%%%%%%%%%%%%%%%%%%%%%%%%%%%%%%%%%%%%%%%%%%%%%%%%%%
\subsection{Operator Power Model} \label{subsec:op_model}
The purpose of the power model is to order the candidate implementations so that \rov~can select the most power efficient design. The e-graph simulation provides us with an average word-level switching activity for each e-class. 

It is a known challenge to accurately estimate operator power consumption based on a word-level RTL implementation as it is highly dependent on downstream transformations and library selection~\cite{Reda2012PowerSurvey}. \rov~mitigates this by encoding certain high-level datapath optimizations, such as arithmetic clustering~\cite{Zimmermann2009DatapathDesign}, in the e-graph~\cite{Coward2022AutomaticE-Graphs}. In this work, we combine the existing theoretical circuit area model from~\cite{Coward2022AutomaticE-Graphs} with the simulated switching activity statistics to estimate the number of two-input gates toggling per clock cycle. For each node $n$ in a given e-class $c$ with child e-classes $c_0,\cdots, c_{k-1}$, we compute a power estimate, $P(n)$.
\begin{equation}
    P(n) = A(n) \times \frac{1}{k+1}\left(T_c + \sum_{i=0}^{k-1} T_{c_i}\right)
\end{equation}
where, $A(n)$ is an estimate of the number of two-input gates required to synthesize that operator, and $T_c$ and $T_{c_i}$ are the operator's output and input toggle frequencies, respectively. For arithmetic operators the area model fixes an architecture, based on known logic synthesis implementations, following the methodology described in~\cite{Coward2022AutomaticE-Graphs}. The area model accounts for operand bitwidths and constant operands. The power model assigns an equal weight to input and output switching activities to approximate the proportion of the gates which transition each cycle. We do not model wire power consumption. In Section~\ref{sec:results} we evaluate how accurately \rov's model is able to estimate power consumption.

As in previous works on RTL optimization using e-graphs, to correctly account for common sub-expressions, we formulate extraction as an integer linear programming (ILP) problem~\cite{Coward2022AutomaticE-Graphs}. Leting $N$ denote the set of e-graph nodes, we use Boolean variables $x_n$ to encode whether we select a particular node $n$ and minimize,
\begin{equation}
    \sum_{n\in N} x_n \times P(n).
\end{equation}
Additional constraints ensure that we extract a valid implementation computing all the module outputs~\cite{Coward2022AutomaticE-Graphs}.

%%%%%%%%%%%%%%%%%%%%%%%%%%%%%%%%%%%%%%%%%%%%%%%%%%%%%%%%%%%%
% RESULTS
%%%%%%%%%%%%%%%%%%%%%%%%%%%%%%%%%%%%%%%%%%%%%%%%%%%%%%%%%%%%
\section{Results} \label{sec:results}

\begin{table*}
    \centering
    \caption{Logic synthesis results comparing the average total power consumption and average area across several delay targets. We compare the baseline, against two implementations generated by \rov, one targeting area optimization and one targeting power optimization. We bold the best result for each metric. We report the relative change vs the baseline and highlight our new contribution in blue. We include the number of nodes in the initial e-graph for each benchmark.}
    \begin{tabular}
    {m{.16\textwidth}
      r 
      r r 
      r r
      r r
     }
        \toprule
        \multirow{2}{*}{Benchmark} & \multirow{2}{*}{Nodes} & \multicolumn{2}{c}{Baseline}      & \multicolumn{2}{c}{Area Optimized}  & \multicolumn{2}{c}{\textcolor{blue}{Power Optimized}} \\
         
         % \cmidrule(lr){12-12}
                  & & Area ($\mu m^2$) & Power ($\mu W$) & Area ($\mu m^2$) & Power ($\mu W$)   & Area ($\mu m^2$) & Power ($\mu W$) \\
        \cmidrule(lr){1-2}
        \cmidrule(lr){3-4}
         \cmidrule(lr){5-6}
         \cmidrule(lr){7-8}
        Comb. Mux Add Tree   & 20 & 32.9 & 	98.2  &       32.8  (\texttt{- 0.4\%}) &   98.8 (\texttt{+ 0.5\%}) &  \best{31.0} (\texttt{- 7.4\%})&  \best{83.2}  (\texttt{-15.5\%})\\
        Address Generation   & 22 & 58.5 &  421.9 & \best{57.1} (\texttt{- 0.2\%}) &  419.2 (\texttt{-0.6\%}) &        57.2  (\texttt{+ 2.2\%})&  \best{301.2} (\texttt{-28.7\%})\\
        Weight Calculation   & 81 & 51.6 & 1141.4 & \best{46.4} (\texttt{-10.2\%}) & 1072.3 (\texttt{-6.1\%}) &        53.3  (\texttt{+ 3.2\%})&  \best{871.5} (\texttt{-23.7\%})\\
        \cmidrule(lr){1-2}
        \cmidrule(lr){3-4}
         \cmidrule(lr){5-6}
         \cmidrule(lr){7-8}
        Pipe. Mux Add Tree~\cite{Munch2000AutomatingDatapaths}& 23 & \best{38.6}  & 852.3 & \best{38.6} (\texttt{\hspace{1.2em}0.0\%}) & 852.3 (\texttt{\hspace{0.7em}0.0\%}) &  44.1 (\texttt{+14.9\%}) &  \best{615.3} (\texttt{-27.2\%})  \\
        Dual Op ALU \cite{Tiwari1998GuardedSynthesis/design}  & 17 &  \best{6.5}  & 186.9 &  \best{6.5} (\texttt{\hspace{1.2em}0.0\%}) & 186.9 (\texttt{\hspace{0.7em}0.0\%}) &  7.5  (\texttt{+15.1\%}) &  \best{146.8} (\texttt{-21.3\%}) \\
        Sequential Reg~\cite{PowerPro2021}                    & 13 &  \best{12.4} & 579.6 & \best{12.4} (\texttt{\hspace{1.2em}0.0\%}) & 579.6 (\texttt{\hspace{0.7em}0.0\%}) &  12.8 (\texttt{+ 2.9\%})&   \best{383.0} (\texttt{-33.9\%})\\
        Dual Path FP Sub~\cite{Coward2023AutomatingE-Graphs}    & 62  & \best{27.8} & 1097.1 & \best{27.8} (\texttt{\hspace{1.2em}0.0\%})& 1097.1 (\texttt{\hspace{0.7em}0.0\%}) & 29.0 (\texttt{+ 3.5\%})& \best{929.4} (\texttt{-14.9\%})\\
        \bottomrule 
    \end{tabular}

    \label{tab:results_table}
\end{table*}

To evaluate \rov's impact on dynamic power consumption, we gathered two sets of benchmarks as shown in Table~\ref{tab:results_table}. The first set of three benchmarks are provided by Intel. The ``Combinational Mux Add Tree'', is taken from Intel low power training materials and comprises of three adders and three muxes. The example demonstrates how the dataflow graph can be rearranged to move particularly high toggling signals towards the outputs, reducing toggling in the rearranged circuit. The second benchmark, ``Address Generation'', is a snippet from production code, which is used as an example of how to perform power optimization in the training materials. It is comprised of two adders, a multiplier and a pair of muxes. The third benchmark, ``Weight Calculation'', is a production two-stage pipelined design computing pixel offsets in the graphics pipeline.

The second set of benchmarks are taken from prior publications~\cite{Tiwari1998GuardedSynthesis/design,Munch2000AutomatingDatapaths,Coward2023AutomatingE-Graphs,PowerPro2021}. The ``Pipelined Mux Add Tree''~\cite{Munch2000AutomatingDatapaths} is similar to the ``Combinational Mux Add Tree'' but introduces a distinct pipelined structure. It is comprised of two adders, three muxes and a pair of registers. The ``Dual Path ALU'' design~\cite{Tiwari1998GuardedSynthesis/design} can optionally perform either a shift or addition. Next, the ``Sequential Reg'' benchmark is used in the PowerPro white paper to demonstrate the tool's sequential clock gating capabilities. It is a combination of registers and a mux. The ``Dual Path FP Sub'' is a pipelined floating point subtractor with a near/far path split~\cite{Coward2023AutomatingE-Graphs}. 

For each design, we pass \rov~the original System Verilog design, \change{which does not contain any existing power optimizations,} along with a json file that specifies the input switching activities. We run \rov~twice generating an area optimized and a power optimized design in System Verilog and an estimate of the power reduction according to \rov's power model. Using a commercial logic synthesis tool targeting a TSMC 5nm cell library, we synthesize the original and \rov~generated designs at a range of delay targets to mitigate the impact of logic synthesis noise~\cite{Coward2022AutomaticE-Graphs}. We provide the commercial tool with the same switching activity configuration as given to \rov. In Table~\ref{tab:results_table}, we show the average circuit area and average total power consumption (including leakage power) reported by the synthesis tool across the range of delay targets. The commercial synthesis tool incorporates a power analysis and optimization tool, which provides relevant power estimates based on the switching activities configured. To ensure the correctness, we verify the cycle-accurate equivalence of the original and \rov~generated designs using a commercial formal equivalence checking tool. 

%%%%%%%%%%%%%%%%%%%%%%%%%%%%%%%%%%%%%%%%%%%%%%%%%%%%%%%%%%%%
% DYNAMIC POWER REDUCTION
%%%%%%%%%%%%%%%%%%%%%%%%%%%%%%%%%%%%%%%%%%%%%%%%%%%%%%%%%%%%
\subsection{Dynamic Power Reduction} \label{subsec:power_reduction_res}
Table~\ref{tab:results_table} compares total power and area results for each of the benchmarks before and after \rov~optimization. \rov~reduces total power consumption by up to \maxpowred~and \avgpowred~on average at the expense of an average \avgareachg~increase in circuit area. The reported power reduction is for a representative set of switching activity configurations. The area optimized designs do not demonstrate the same power reduction but show some limited area improvements. For several designs the area optimization could not find any improvement, returning the baseline implementation. Whilst \rov~only models dynamic power, we evaluate based on total power, including leakage power. 

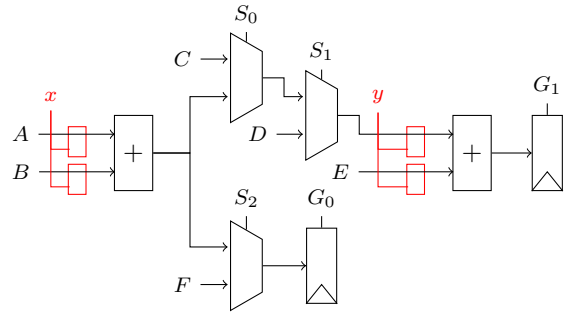
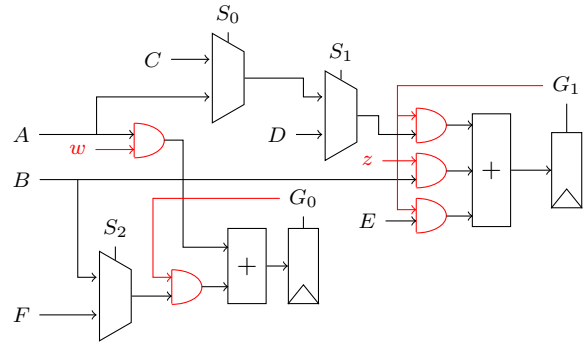
\begin{figure}
    \centering
    \subfloat[Baseline design (black). In~\cite{Munch2000AutomatingDatapaths} the authors add the transparent registers (red), where $x=(\overline{S_2}\& G_0)\| (S_0 \& \overline{S_1} \& G_1)$ and $y=G_1$.] {\begin{tikzpicture}

	% Place nodes using a matrix
    \node at (0,0.25) (a)  {\footnotesize $A$};
    \node at (0,-0.25) (b) {\footnotesize $B$};
    
    \node[red] at (0.4,0.75) (x) {\footnotesize $x$};
    \node[red] at (4.75,0.75) (y) {\footnotesize $y$};
    
    \node at (2.15,1.25) (c) {\footnotesize $C$};
    \node at (3.15,0.25) (d) {\footnotesize $D$};
    \node at (4.25,-0.25) (e) {\footnotesize $E$};
    \node at (2.15,-1.75) (f) {\footnotesize $F$};
    \node at (3,1.85) (s0) {\footnotesize $S_0$};
    \node at (4,1.35) (s1) {\footnotesize $S_1$};
    \node at (3,-0.6) (s2) {\footnotesize $S_2$};
    \node at (4,-0.6) (g0) {\footnotesize $G_0$};
    \node at (7,0.9) (g1) {\footnotesize $G_1$};
    
    \node[draw=red,shape=rectangle, minimum height=0.4cm] at (0.75, 0.15) (latch1) {};
    \node[draw=red,shape=rectangle, minimum height=0.4cm] at (0.75,-0.35) (latch2) {};
    \node[draw,shape=rectangle, minimum height=1cm] at (1.5,0) (add1) {$+$};
    \node[draw,shape=trapezium,inner xsep=10pt, inner ysep=6pt,rotate=270] at (3,1) (sels0) {};
    \node[draw,shape=trapezium,inner xsep=10pt, inner ysep=6pt,rotate=270] at (4,0.5) (sels1) {};
    \node[draw=red,shape=rectangle, minimum height=0.4cm] at (5.25,0.15) (latch1) {};
    \node[draw=red,shape=rectangle, minimum height=0.4cm] at (5.25,-0.35) (latch2) {};
    \node[draw,shape=rectangle, minimum height=1cm] at (6,0) (add2) {$+$};
    \node[draw,shape=rectangle, minimum height=1cm, minimum width=0.4cm] at (7,0) (regg1) {};
    
    \node[draw,shape=trapezium,inner xsep=10pt, inner ysep=6pt,rotate=270] at (3,-1.5) (sels2) {};
    \node[draw,shape=rectangle, minimum height=1cm, minimum width=0.4cm] at (4,-1.5) (regg0) {};

    \draw[->] (a) -> (1.25,0.25);
    \draw[->] (b) -> (1.25,-0.25);
    \draw[->] (add1) -- (2.25,0) -- (2.25,0.75) -- (2.75,0.75);
    \draw[->] (add1) -- (2.25,0) -- (2.25,-1.25) -- (2.75,-1.25);
    \draw[->] (sels0) -- (3.5,1) |- (3.75,0.75);
    \draw[->] (sels1) -- (4.5,0.5) |- (5.75,0.25);
    \draw[->] (sels2) -- (regg0);
    \draw[->] (add2) -- (regg1);
    \draw[->] (c) -- (2.75,1.25);
    \draw[->] (d) -- (3.75,0.25);
    \draw[->] (e) -- (5.75,-0.25);
    \draw[->] (f) -- (2.75,-1.75);
    \draw[] (s0) -- (sels0);
    \draw[] (s1) -- (sels1);
    \draw[] (s2) -- (sels2);
    \draw[] (g0) -- (regg0);
    \draw[] (g1) -- (regg1);
    
    \draw[red] (x) |- (0.65,0.05);
    \draw[red] (x) |- (0.65,-0.45);

    \draw[red] (y) |- (5.15,0.05);
    \draw[red] (y) |- (5.15,-0.45);
    
    \draw[] (regg0.south east) -- (4,-1.75) -- (regg0.south west); 
    \draw[] (regg1.south east) -- (7,-0.25) -- (regg1.south west); 

 \end{tikzpicture}\label{fig:orig_pipe_mux_tree}}
    \qquad
    \subfloat[\rov~generated design. \rov~rearranged the mux tree and added data gating (red), where $w=\overline{S_2}\&G_0$ and $z=G_1\& \overline{S_1}\& S_0$. ] {\begin{tikzpicture}

	% Place nodes using a matrix
    \node at (-0.25,0.25) (a)  {\footnotesize $A$};
    \node at (-0.25,-0.35) (b) {\footnotesize $B$};
    
    % \node[red] at (0.4,0.75) (x) {\footnotesize $x$};
    % \node[red] at (4.75,0.75) (y) {\footnotesize $y$};
    
    \node at (1.5,1.25) (c) {\footnotesize $C$};
    \node at (3.15,0.25) (d) {\footnotesize $D$};
    \node at (4.35,-0.9) (e) {\footnotesize $E$};
    \node[red] at (4.35,-0.1) (z) {\footnotesize $z$};
    \node[red] at (0.5,0.05) (w) {\footnotesize $w$};
    \node at (-0.25,-2.15) (f) {\footnotesize $F$};
    \node at (2.5,1.85) (s0) {\footnotesize $S_0$};
    \node at (4,1.35) (s1) {\footnotesize $S_1$};
    \node at (1,-1) (s2) {\footnotesize $S_2$};
    \node at (3.5,-0.6) (g0) {\footnotesize $G_0$};
    \node at (7,0.9) (g1) {\footnotesize $G_1$};
    
    % \node[draw=red,shape=rectangle, minimum height=0.4cm] at (0.75, 0.15) (latch1) {};
    % \node[draw=red,shape=rectangle, minimum height=0.4cm] at (0.75,-0.35) (latch2) {};
    \node[draw,shape=rectangle, minimum height=1cm] at (2.75,-1.5) (add1) {$+$};
    \node[draw,shape=trapezium,inner xsep=10pt, inner ysep=6pt,rotate=270] at (2.5,1) (sels0) {};
    \node[draw,shape=trapezium,inner xsep=10pt, inner ysep=6pt,rotate=270] at (4,0.5) (sels1) {};

    \draw[red] (5,0.6) arc(90:-90:0.4cm and 0.23cm);
    \draw[red] (5,0.15) -> (5,0.6);

    \draw[red] (5,0) arc(90:-90:0.4cm and 0.23cm);
    \draw[red] (5,-0.45) -> (5,0);

    \draw[red] (5,-0.6) arc(90:-90:0.4cm and 0.23cm);
    \draw[red] (5,-1.05) -> (5,-0.6);
    \node[draw,shape=rectangle, minimum height=1.5cm] at (6,-0.225) (add2) {$+$};
    \node[draw,shape=rectangle, minimum height=1cm, minimum width=0.4cm] at (7,-0.225) (regg1) {};
    
    \node[draw,shape=trapezium,inner xsep=10pt, inner ysep=6pt,rotate=270] at (1,-1.9) (sels2) {};
    \node[draw,shape=rectangle, minimum height=1cm, minimum width=0.4cm] at (3.5,-1.5) (regg0) {};

    \draw[red] (1.75,-1.55) arc(90:-90:0.4cm and 0.23cm);
    \draw[red] (1.75,-2) -> (1.75,-1.55);
    
    % AND GATE 1
    \draw[red] (1.25,0.4) arc(90:-90:0.4cm and 0.23cm);
    \draw[red] (1.25,0.4) -> (1.25,-0.05);
    \draw[->] (1.65,0.175) -- (1.9,0.175) -- (1.9,-1.25)-- (2.5,-1.25);
    
    \draw[->] (0.5,-0.35) -- (0.5,-1.65) -- (0.75,-1.65);
    \draw[->] (b) -- (5,-0.35);
    \draw[->,red] (z) -- (5,-0.1);
    \draw[->] (e) -- (5,-0.9);

    \draw[->] (5.4,-0.835) -- (5.75,-0.835);
    \draw[->] (5.4,-0.235) -- (5.75,-0.235);
    \draw[->] (5.4,0.375) -- (5.75,0.375);
    
    \draw[->] (a) -- (0.75,0.25) -- (0.75,0.75) -- (2.25,0.75);
    \draw[->] (0.75,0.25) -> (1.25,0.25);
    \draw[red,->] (w) -> (1.25,0.05);
    
    \draw[red,->] (g1) -- (4.75,0.9) -- (4.75,0.5) -- (5,0.5);
    \draw[red,->] (4.75,0.5) -- (4.75,-0.75) -- (5,-0.75);

    \draw[->] (sels0) -- (3.5,1) |- (3.75,0.75);
    \draw[->] (sels1) -- (4.5,0.5) |- (5,0.25);
    \draw[->] (sels2) -- (1.75,-1.9);
    
    \draw[->] (2.15,-1.775) -- (2.5,-1.775);
    
    \draw[->] (add1) -- (regg0);
    \draw[->] (add2) -- (regg1);
    \draw[->] (c) -- (2.25,1.25);
    \draw[->] (d) -- (3.75,0.25);
    
    \draw[->] (f) -- (0.75,-2.15);
    \draw[] (s0) -- (sels0);
    \draw[] (s1) -- (sels1);
    \draw[] (s2) -- (sels2);
    \draw[] (g0) -- (regg0);
    \draw[] (g1) -- (regg1);

    \draw[->,red] (g0) -- (1.5,-0.6) -- (1.5,-1.65) -- (1.75,-1.65);

    % \draw[red] (x) |- (0.65,0.05);
    % \draw[red] (x) |- (0.65,-0.45);

    % \draw[red] (y) |- (5.15,0.05);
    % \draw[red] (y) |- (5.15,-0.45);
    
    \draw[] (regg0.south east) -- (3.5,-1.75) -- (regg0.south west); 
    \draw[] (regg1.south east) -- (7,-0.475) -- (regg1.south west); 

 \end{tikzpicture}}
    \caption{Circuit diagrams of the ``Pipelined Mux Add Tree'' benchmark with power optimizations from prior work and from \rov.
    }
    \label{fig:pipe_mux_tree}
\end{figure}

To best understand the benefit of exploring arithmetic, area and power in tandem, we study an open-source benchmark. Figure~\ref{fig:pipe_mux_tree} shows the circuits corresponding to the baseline ``Pipelined Mux Add Tree'', the design proposed in~\cite{Munch2000AutomatingDatapaths} and the \rov~generated version. The optimizations proposed in~\cite{Munch2000AutomatingDatapaths} add transparent registers to both adder inputs, as this work only added operators. Meanwhile, \rov~performs an entirely different optimization, re-ordering the dataflow graph to push the adders towards the output of the circuit. Note that the \rov~generated design contains a three input adder, which, thanks to \rov's comprehension of logic synthesis optimizations, is recognised as only a single carry-save adder, rather than two full carry-propagate adders. \rov~then inserts area efficient data gating on the adder inputs to save power. Synthesizing the design proposed in~\cite{Munch2000AutomatingDatapaths}, the \rov~generated architecture is strictly better, consuming 11\% less power within 17\% less area. 

For the ``Combinational Mux Add Tree'' \rov~once again re-orders the mux tree converting three separate adders to one single adder taking four inputs. This differs from the solution proposed in the Intel training materials. \rov's design reduces both power and area by 10\% when compared to the design proposed in the training materials. In the ``Address Generation'' benchmark, \rov~deploys data gating as recommended by the training material, but also an optimization to combine two adders into one three input adder. This area optimization offsets the overhead of the gating operators, leading to only a 2.2\% increase in area. For the ``Dual Op ALU'' and ``Sequential Reg'' benchmarks, \rov~is able to rediscover the optimizations proposed in~\cite{Tiwari1998GuardedSynthesis/design} and~\cite{PowerPro2021}, demonstrating \rov's ability to generalize prior work. Lastly, \rov~recognises the distinct computational paths in the ``Dual Path FP Sub'' and inserts the appropriate clock gating for each path.

For all but two benchmarks \rov~ran in less than 10 seconds, taking only a few seconds for the majority. For the ``Address Generation'' and ``Weight Calculation'' benchmarks \rov~ran for 130 seconds and 160 seconds, respectively. These long running cases were dominated by the ILP solver. Comparing the reduction in power consumption predicted by \rov's model against the actual impact reported by logic synthesis, we see that for a group of five benchmarks the model provides a relevant estimate of the power reduction, within 14 percentage points of the actual. However, for the ``Combinational Mux Add Tree'' and ``Address Generation'' the model overestimates the improvement in power consumption by around 45 percentage points. We attribute this to two causes. First, for both benchmarks, the area model predicted an area reduction that was not realized. Second, the power model uses only a simple linear relationship between operator power and toggle frequencies. 
%%%%%%%%%%%%%%%%%%%%%%%%%%%%%%%%%%%%%%%%%%%%%%%%%%%%%%%%%%%%
% DATA DEPENDENT DESIGN
%%%%%%%%%%%%%%%%%%%%%%%%%%%%%%%%%%%%%%%%%%%%%%%%%%%%%%%%%%%%
\subsection{Data Dependent Design}\label{subsec:data_dependent_design}
To demonstrate how \rov~is capable of tailoring the implementation to the computation, we study how \rov's output changes as we modify the switching activities of \change{mux select and register enable signals}. We focus on the ``Pipelined Mux Add Tree'' as shown in Figure~\ref{fig:pipe_mux_tree}. In Table~\ref{tab:data_depend_design} \change{we pass \rov~four different switching activity configurations and report the optimizations selected by \rov~for each}. \change{Given Cfg.~1, \rov~elects to insert data gating using the $S_0,S_1,S_2$ and $G_1$ signals. Given Cfg.~3, where we increase the switching activity for all signals, \rov~instead elects to insert a single transparent register.} In Cfg. 4, we see \rov~use the $G_0$ signal to data gate, which we do not see in other configurations. The final columns show how the power benefit of \rov's optimizations varies with switching activities.

\begin{table}[]
    \centering
    \setlength\extrarowheight{2pt}
    \caption{Each row represents a different switching activity configuration (Cfg.) for the mux select and register enable signals in the ``Pipelined Mux Add Tree'' (Figure~\ref{fig:orig_pipe_mux_tree}). For each Cfg., if \rov~inserted a data gate (transparent register) using one of these signals, we color the corresponding cell green (purple).}
    \begin{tabular}{cccccccc}
         \toprule
          & \multicolumn{3}{c}{Muxes} & \multicolumn{2}{c}{Registers} & \multicolumn{2}{c}{Total Power (mW)}\\
               
         \cmidrule(lr){2-4}
         \cmidrule(lr){5-6}
         \cmidrule(lr){7-8}
          Cfg. & $S_0$ & $S_1$ & $S_2$ & $G_0$ & $G_1$           & Baseline    & ROVER  \\
         % \cmidrule(lr){1-1}
         % \cmidrule(lr){2-4}
         \cmidrule(lr){1-6}
         \cmidrule(lr){7-8}
         % \midrule
         1 & \ccp 0.1 & \ccp 0.1 & \ccp 0.1 & 0.1 & \ccp 0.1      & 1.09 & 0.76 (-30\%)\\
         
         2 & \ccp 0.1 & \ccp 0.1 & \ccp 0.1 & 0.8 & \ccg 0.8      & 1.09 & 0.95 (-14\%)\\
         
         3 & 0.8      & 0.8      & 0.8   & 0.8    & \ccg 0.8      & 1.30 & 1.15 (-11\%) \\
         
         4 & \ccp 0.8 & \ccp 0.8 & \ccp 0.8 & \ccp 0.1 & \ccp 0.1 & 1.29 & 1.03 (-20\%) \\
         \bottomrule
    \end{tabular}
    \label{tab:data_depend_design}
\end{table}

%%%%%%%%%%%%%%%%%%%%%%%%%%%%%%%%%%%%%%%%%%%%%%%%%%%%%%%%%%%%
% CONCLUSION
%%%%%%%%%%%%%%%%%%%%%%%%%%%%%%%%%%%%%%%%%%%%%%%%%%%%%%%%%%%%
\section{Conclusions and Future Work}\label{sec:conclusion}
This paper describes how to encode power optimizations, such as operand isolation and clock gating, as local equivalence preserving rewrites. By phrasing power reduction as a rewrite problem we can combine it with existing arithmetic rewrites to explore both power and area in tandem. We developed an e-graph based rewriting framework, \rov, that can simultaneously explore and balance the area-power tradeoff. The e-graph enables efficient simulation of many functionally equivalent implementations as we only need to simulate one node from each class. Optimizing a set of benchmarks using \rov~we see a \avgpowred~reduction in total power consumption on average for just an average circuit area increase of~\avgareachg. We show the importance of \rov's understanding of downstream logic synthesis optimizations, leading to designs not seen in prior work.

Whilst \rov~is able to generalize the majority of optimizations discussed in prior work, the optimization presented in Figure~\ref{fig:orig_pipe_mux_tree}, is not expressible via the current set of rewrites. We can describe this as a resource sharing problem, a general optimization that future work on e-graph rewriting will address. We will also automatically derive useful case-splits from simulation stimuli to insert into the RTL. 

% \begin{figure*}
% \input{transp_reg}    
% \end{figure*}

\bibliographystyle{IEEEtran}
\bibliography{references}

\end{document}